\documentstyle[12pt]{article}
\def\be{\begin{equation}}
\def\ee{\end{equation}}
\def\bea{\begin{eqnarray}}
\def\eea{\end{eqnarray}}

\author{Hans - J\"urgen Schmidt}

\title{Surface Layers in General Relativity and Their
Relation to Surface Tensions}

\date{}
\begin{document}
\maketitle

\centerline{Universit\"at Potsdam, Institut f\"ur Mathematik, Am
Neuen Palais 10} 
 \centerline{D-14469~Potsdam, Germany,  E-mail:
 hjschmi@rz.uni-potsdam.de}

\begin{abstract}
For a thin shell, the intrinsic 3-pressure will be shown to be 
analogous  to $ -A$, where $A$ is the classical 
 surface tension: First, interior and exterior Schwarzschild
 solutions will be matched together such that the surface 
layer generated at the common boundary has no gravitational 
mass; then its intrinsic 3-pressure represents a surface tension 
fulfilling Kelvin's relation between mean curvature and pressure 
difference in the Newtonian limit. Second, after a suitable definition 
of mean curvature, the general relativistic analogue  to Kelvin's relation 
will be proven to be contained in the equation of motion of the 
surface layer.
\end{abstract}

\section{Introduction} 

In general relativity, an energy-momentum tensor concentrated
 on a time-like hypersurface is called a surface layer. Via Einstein's 
equations it is related to nonspurious jumps of the 
Christoffel affinities or equivalently to jumps of the second 
fundamental tensor. In [1-6] there have been given algorithms 
for their calculating, and in [1], [4], and [7-13] the spherically
 symmetric case was of a special interest. The discussion of 
disklike layers as models for accretion disks was initiated in 
[14] and [15], and [21] contains surface layers  as sources 
of the Kerr geometry.

A surface layer has no component in the normal direction, 
otherwise the delta-like character would be destroyed. Hence, an 
ideal fluid with nonvanishing pressure cannot be concentrated on an 
arbitrary thin region and therefore in most cases, the layer is 
considered to be composed of dust. Additionally, an intrinsic 
3-pressure is taken into account in [16] (cf. [17]), where for
 a spherically symmetric configuration each shell 
$r =$ const is thought to be composed of  identical particles
 moving on circular orbits without a preferred direction, and the 
tangential pressure ($=$ intrinsic 3-pressure) is due to particle
 collisions, whereas 
a radial pressure
does not appear. (See also [10], [13], and [26],  where the equation of
 motion for a spherically
symmetric layer has been discussed.)

Such an intrinsic 3-pressure as 
well as surface 
tensions are both of the physical dimension
``force per unit length." There 
the question arises whether an intrinsic 3-pressure of a surface
layer may be related to a surface tension, 
and the present paper deals with just this question. Then
a general relativistic formulation of 
thermodynamics (cf. [18] or [19] for this and [20] for
nonrelativistic surface tensions) can be 
completed by equations for surface tensions to answer,
e.g., the question how long 
a drop, say, a liquid comet, remains connected while
 falling
 towards a compact object. The only paper 
concerned with such questions seems to be [22]. There
the influence of surface tensions on the propagation 
of gravitational waves has been calculated
by perturbation methods, yielding a possibly measurable effect.

The paper proceeds as follows: 
Section 2 contains Kelvin's relation for nonrelativistic surface
tensions. Section 3 discusses the 
matching of the interior to the exterior Schwarzschild solution
such that a surface tension appears (calculations are 
found in the appendices) and compares with
Kelvin's relation in the Newtonian limit. 
Sections 4 and 5 are devoted to the nonspherically
symmetric case. Section 4 contains 
a suitable definition of mean curvature and Section 5
 deduces Kelvin's relation  from the 
equation of motion of the surface layer without any weak
 field assumptions.

\section{Nonrelativistic  Surface Tensions}

Imagine a drop of some liquid moving in vacuo. 
Its equilibrium configuration is a spherical
one, and Kelvin's relation [23] 
between surface tension $A$, 
pressure difference (outer minus inner
pressure) $\Delta P$, and 
mean curvature $H$ ($H = 1/R$  for a sphere of radius $R$)
 reads
\be
     \Delta P = -2H A \, .
\ee
$A$  is a material-dependent constant. 
Equation (1) means, an energy $A \cdot \Delta F$
 is needed to increase the surface area by $\Delta F$.
This supports 
our description of $-A$  as a kind of intrinsic pressure. 
(But of course, it is a quite 
different physical process: Pressure, say, of an ideal gas, can be explained
 by collisions of freely 
moving particles, whereas the microphysical explanation of surface
tensions requires the determination 
of the intermolecular potential, which looks like
$$
\Phi(r) = - \mu r^{-6}  +N e^{-r/\rho}
$$
with certain constants $\mu$, $N$, 
and $\rho$; see [20]. 
In this context the surface has a thickness of about
$10^{-7}$ cm but in most cases this thickness may be neglected.)

In addition, for a nonspherically symmetric 
surface, the mean curvature $H$  in equation (1) may
be obtained from the principal curvature 
radii   $R_1$, $R_2$ 
by means of the relation
\be
H = \frac{1}{2R_1}  + \frac{1}{2R_2}   \, .
\ee

\section{Spherical Symmetry}

Now the drop shall be composed of 
an incompressible liquid. For a general relativistic
description we have to take the interior Schwarzschild  solution
\bea
ds^2 = -   \left[ \frac{3}{2}(1-r_g/r_0)^{1/2}
-  \frac{1}{2}  ( 1-r_gr^2/r_0^3) ^{1/2}
\right]^2 
dt^2    \nonumber \\
+  \frac{dr^2}{1-r_gr^2/r_0^3}
+  r^2   d\Omega^2    \\
d\Omega^2 = d\psi^2 + \sin^2 \psi d \varphi^2   \nonumber
\eea
whose energy-momentum tensor 
represents ideal fluid with energy density $\mu = 
 3 r_g / \kappa  r_0^3$  and  pressure
\be
p(r) = \mu \cdot 
\frac{   ( 1-r_gr^2/r_0^3) ^{1/2} - 
(1-r_g/r_0)^{1/2}
}{3 (1-r_g/r_0)^{1/2} -
   ( 1-r_gr^2/r_0^3) ^{1/2}} \, .
\ee
One has $p(r_0)= 0$, and therefore 
usually $0 \le r \le r_0$  is considered. But now we require only a
nonvanishing pressure at the inner 
surface and take (3) for values $r$ with $0 \le r \le R$ 
 and a fixed $R <r_0$ only. The gravitational mass 
of the inner region equals
\be
     M= \mu \cdot 4 \pi  R^3/3 = r_g(R/r_0)^3/2 
\ee
where $G= c = 1$. The outer region shall be empty and
therefore we have to  insert the Schwarzschild
solution for $r \ge R$. Neglecting the gravitational mass\footnote{The
 vanishing of the  gravitational mass 
is required to single out the properties of an intrinsic
3-pressure. In general, a 
surface layer is  composed of  both parts, not at least to ensure the
 validity of the energy 
condition $T_{00} \ge \vert T_{ik} \vert $
 which holds for all known types of matter.}
 of the boundary $ \Sigma \subset  V_4$  which is defined
by $r - R =  0$,  just $M$ of equation (5) has 
to be  used as the mass parameter of the exterior
Schwarzschild metric. Then only delta-like tensions 
appear at $\Sigma $,  and $\Delta P = - p(R) <0$. On 
$\Sigma $, the
energy-momentum tensor is
\be
T_i^k = \tau_i^k \cdot \delta_\Sigma 
\ee
the  nonvanishing components of which are
\be
\tau_2^2 = \tau_3^3 =-p(R)R/2(1 - 2M/R)^{1/2}
\ee
cf. Appendix A.

The $ \delta_\Sigma $ distribution is defined such 
that for all smooth scalar functions $f$  the invariant integrals
$$
\int_\Sigma \ 
f \ d^{(3)} x \qquad {\rm  and}  \qquad 
\int_{V_4} \ 
f \cdot     \delta_\Sigma    d^{(4)} x
$$
coincide. For a more detailed 
discussion of distribution-valued tensors
 in curved spacetime, cf. [24].

The fact that all components $\tau_0^\alpha$
 in (6) vanish 
reflects the  nonexistence of a delta-like
gravitational mass on $ \Sigma $.

Now consider the Newtonian  limit $M/R \ll 1$; then 
the mean curvature becomes again $H= 1/R$,
and together with (7) Kelvin's relation (1) is just 
equivalent to
\be
\tau_2^2 = \tau_3^3 = - A [ 1 + O(M/R) ] \, .
\ee
Therefore: At least for static spherically symmetric 
configurations and weak fields a delta-like
negative tangential pressure coincides 
with the classical surface tension.

In the next two sections we investigate 
to what extent these presumptions are necessary.

\section{Mean Curvature in Curved Space-Time}

To obtain a general relativistic analogue   to Kelvin's
 relation (1) we have to define the mean
curvature $H$ of the time-like hypersurface 
$\Sigma$ contained in a space-time $V_4$ such that for weak
fields just the usual mean curvature arises.

To get the configuration  we have in mind, we make the ansatz
\be
 \tau_{\alpha \beta} = - A ( g_{\alpha \beta}
 + u_\alpha u_\beta ) \, \qquad u_\alpha u^\alpha = -1
\ee
with $A >0$.  Thereby again the gravitational 
mass of $ \Sigma$  will be neglected. Now the mean curvature
shall be defined. But there is a problem: 
In general, there does not exist a surface $ S \subset 
\Sigma$  which can
serve as ``boundary at a fixed moment" for which 
we are  to determine the mean curvature. To
circumvent this problem we start considering the special case
\be
    u_{\alpha \| \beta}  =       u_{\beta \|  \alpha }   \,     .
\ee
Then there exists a scalar 
$t$ on $\Sigma$  such that 
$u_\alpha  = t_{\| \alpha}$, 
and the surface $S \subset  \Sigma$  defined by $t =0$
 may be
called ``boundary at a fixed moment." 
Now $S$ has to be embedded into a ``space at a fixed
moment'': We take intervals of geodesics starting 
from  points of $S$  in the normal direction $n_i$  and
the opposite one. The union $V_3$ of these geodetic 
segments will be called ``space at a fixed moment," and $S \subset V_3$
 is simply a two-surface in a three-dimensional positive 
definite Riemannian manifold, for which
mean curvature has a definite sense: 
Let $v^\alpha$, $w^\alpha$
 be the principal curvature directions inside $S$  and
 $R_1$, $R_2$ the corresponding principal 
curvature radii, then equation (2) applies to obtain $H$.

Of course,   $v^\alpha   w_\alpha =0 $  holds, 
and $ v^\alpha   v_\alpha
=   w^\alpha   w_\alpha  = 1$ shall be attained. Then, inserting the second
fundamental tensor (cf. Appendix B), this becomes equivalent to
\be
H = \frac{1}{2} (v^\alpha   v^\alpha
+   w^\alpha   w^ \alpha    )
k_{\alpha \beta }
= \frac{1}{2} (g^{\alpha \beta} + u^\alpha u^\beta)
k_{\alpha \beta }
\, . 
\ee
But this latter relation makes sense without 
any reference to condition (10). Therefore, we define
(11) to  be the general relativistic 
analogue  to the mean curvature of a surface. For the case   
$ H^+ \ne H^- $
we take their arithmetic mean\footnote{This
 choice can  be accepted noting that  
in the Newtonian limit 
$$
 \vert H^+ - H^- \vert \ll  \vert H^+ + H^- \vert  
$$
anyhow.}
$$
 H = (    H^+ + H^- )/2   \,   ,
$$
 and  then we obtain from  (9) and (11)
\be
     -2HA = \frac{1}{2} \left(
k_{\alpha \beta }^+ + k_{\alpha \beta }^- \right)
 \tau^{\alpha \beta }     \, . 
\ee
To compare this with Kelvin's relation 
we have to relate the right-hand side of equation (12) to
the pressure difference $\Delta P$ at $\Sigma$. 
To this end 
we investigate the equation of motion for the surface
layer.

\section{Equation of Motion for the Surface Layer}

The equation of  motion, $T^k_{i;k}=  0$, contains products 
of $\delta $ distributions and
$\theta$-step functions\footnote{
$\theta(x) =1 $ for $x \ge 0$, and $\theta(x) =0 $ else.}
  at points 
where $ \Gamma^i_{jk}$
  has a  jump discontinuity. These products
require special care; 
cf. [25] for a discussion of  his point. 
But defining   $ \theta \cdot \delta = \frac{1}{2} \delta $
 we obtain (cf.  Appendix C)
\bea
\Delta P \equiv \Delta n_i n^k T^i_k =  \frac{1}{2} \left(
k_{\alpha \beta }^+ + k_{\alpha \beta }^- \right)
 \tau^{\alpha \beta }
 \qquad {\rm   and }  \\
  \Delta T^1_\alpha \equiv \Delta n_i e^k_\alpha 
 T^i_k =  - \tau^\beta_{\alpha \| \beta} \, .
\eea
From equation (14) we  see  the following: 
The equation $ \tau^\beta_{\alpha \| \beta} =0$ 
holds only under the
additional presumption that the regular 
(i.e., not delta-like) part of $ T^1_\alpha$ 
has no jump on $\Sigma$. This
condition is fulfilled e.g.,  presuming $\Sigma$
 to be such a boundary that the regular energy  flow does
not cross it and the four-velocity is parallel to $\Sigma$
  in both $V_+$  and $V_-$. This we will presume in the
following. Then $\Delta P$
 is indeed 
the difference of the pressures on both sides, and together with
equations (12) and (13) 
we obtain exactly Kelvin's  relation (1). That means, it is the definition
of mean curvature used here that enables 
 us to generalize Kelvin's formula to  general relativity.
Furthermore, $O(M/R)$  of equation (8) vanishes.

Finally we want to discuss 
the equation $ \tau^\beta_{\alpha \| \beta} =0$. 
Transvection with $u^\alpha$  and 
$ \delta^\alpha_\gamma + u^\alpha u_\gamma $
 yields
\be
 u^\alpha_{\| \alpha} =0 
\qquad {\rm and} \qquad
 u^\alpha u_{\gamma \| \alpha}
+
(\ln A)_{\| \alpha}
( \delta^\alpha_\gamma + u^\alpha u_\gamma )
=0
\ee
respectively. But $A$ is a 
constant here, and therefore $u^\alpha$
is an expansion-free  geodesic vector
field in $\Sigma$.\footnote{But observe that the  $u^\alpha$ 
lines are geodesics in $V_4$  under additional presumptions only.}

Here, we have only considered a phenomenological theory of 
surface tensions, and, of course,
a more detailed theory has to include 
intermolecular forces. But on that phenomenological level
equations (9) and (15) together with 
Kelvin's relation (1) (which has been shown to follow from
the equation of motion) and 
$A =$ const as a (solely temperature-dependent) equation of state 
complete the usual general relativistic 
Cauchy problem for a thermodynamical system by including
surface tensions.

\section*{Appendix A}

To deduce equation (7) we take 
proper time $\xi^0$  and angular coordinates 
 $\psi = \xi^2$  and $\varphi  = \xi^3$.
Then, inside $\Sigma$,
$$
ds^2 = - \left( d \xi^0 \right)^2  + R^2 d\Omega^2    \, .
$$
Using equation (20) and the exterior Schwarzschild
 solution one obtains 
$k_{\alpha \beta }^+ $
 the nonvanishing components of which are
$$
k_{00}^+
 = - M/R^2(l - 2M/R)^{1/2} \, , \qquad
   k_{22}^+ =R(1 - 2M/R)^{1/2} 
$$
and
$$
k_{33}^+ =  k_{22}^+
   \sin^2 \psi
$$
because of spherical symmetry. To 
avoid long calculations with the metric (3) one can proceed
as follows. By construction, $\tau_{00} =0$,  and 
together with equation (21) and the spherical
symmetry    
$  k_{22}^- =  k_{22}^+$, $k_{33}^- = k_{33}^+$
follows. 
From equation (24), equation (7) follows then
immediately without the necessity 
of determining the actual value  of $  k_{00}^-  $.

And to be independent of the discussions 
connected with equation (23), we deduce equation (24)
another way. First, (independent of  surface 
layers), for an arbitrary timelike hypersurface and a
coordinate system such that (22) holds, we have
\be
\kappa  T_{11} = \frac{1}{2}
\left(
{}^{(3)}R + k^2 -k_{\alpha \beta }k^{\alpha \beta }
\right)
\ee
where $\  {}^{(3)}R$  is the curvature scalar 
within that surface. Now turn to a surface layer with 
$ k^+_{\alpha \beta } \ne  k^-_{\alpha \beta } $. 
Then equation (16) splits into a ``$+$''  and a ``$-$''  equation,
having in common solely   $\  {}^{(3)}R$. 
Inserting all this into equation (21), one obtains
$$
  \frac{1}{2} \left(
k_{\alpha \beta }^+ + k_{\alpha \beta }^- \right)
 \tau^{\alpha \beta } =
 \frac{1}{2 \kappa }  \left(
k_{\alpha \beta }^+ + k_{\alpha \beta }^- \right)
 \left(
g^{\alpha \beta } \Delta k - \Delta  k^{\alpha \beta }
\right)
$$
$$
= \frac{1}{2 \kappa } \left[
(k^+)^2 - (k^-)^2 
- k_{\alpha \beta }^+ k^{+ \alpha \beta }
+ k_{\alpha \beta }^-  k^{- \alpha \beta } 
\right]
= T^+_{11} - T^-_{11}
$$
i.e., just equation (24) below.

\section*{Appendix B}

To make the paper more readable, some 
conventions and formulas shall be given.  Let $\xi^\alpha$,  
 $\alpha  =
0,2,3$, be coordinates in $\Sigma $
and $x^i$, $i=0,1,2,3$, those for $V_4$. 
The embedding $\Sigma \subset V_4$  is performed by
functions $x^i(\xi^\alpha)$
 whose derivatives
\be
e^i_\alpha = \partial x^i/\partial \xi^\alpha \equiv x^i_{, \alpha}
\ee
form a triad field in $\Sigma$.  $\Sigma$ 
  divides,  at least 
locally, $V_4$  into two connected  components, 
$V_+$  and $V_-$, and the normal $n_i$, defined by
\be
 n_i n^i =1\, , \qquad n_i e^i_\alpha =0 
\ee
is chosen into the $V_+$ direction 
(which can be thought being the outer region).  Possibly $V_+$
  and $V_-$  are endowed with different 
coordinates $x^i_+$, $x^i_-$  and metrics
 $g_{ik^+}$
and  $g_{ik^-}$, respectively. For
this case all subsequent formulas 
had to be indexed with $+/-$, and only the inner 
metric of $\Sigma$, its
first fundamental tensor
\be
g_{\alpha \beta} = e^i_\alpha e^k_\beta \  g_{ik}
\ee
has to be the same in both cases. 
As usual, we require $g_{ik}$ 
to be $C^2$-differentiable except for
jumps of $g_{ij,k}$  at $\Sigma$. Covariant derivatives 
within $V_4$  and $\Sigma$  will be denoted by $ \  ; \ $  and 
 $\  \|  \ $  respectively.

The second fundamental tensor 
$k^\pm_{\alpha \beta}$
 on both sides of $\Sigma$  is defined by
\be
k^\pm_{\alpha \beta}=
\left( e^i_\alpha e^k_\beta n_{i;k}     \right)^\pm
=
\left( e^i_{\alpha;k} e^k_\beta n_{i}     \right)^\pm
\ee
and the difference, 
$ \Delta k_{\alpha \beta} = k^+{\alpha \beta} -
k^-_{\alpha \beta}$, 
$\Delta k = g^{\alpha \beta } \Delta k_{\alpha \beta}$, 
enters the energy-momentum tensor via
equation  (6) and the relation
\be
\tau^{ik} \equiv e^i_\alpha e^k_\beta 
\tau^{\alpha \beta} \, , 
 \qquad {\rm  where} \qquad  \kappa \tau_{\alpha \beta} = 
g_{\alpha \beta}\Delta  k - \Delta   k_{\alpha \beta}
\ee
cf. e.g., [4]. 
From equation (18) and the  Lanczos equation (21) one obtains 
$n^i\tau_{ik} = 0$, i.e., indeed the absence of a delta-like
 energy flow in the normal direction.

\section*{Appendix C}

Now take a special coordinate 
system: $x^\alpha  = \xi^\alpha$, and the 
$x^1$ lines are geodesics starting from $\Sigma$ 
into $n_i$ direction with natural parameter $x^1$. 
Then the  line element of $V_4$ reads
\be
ds^2 =  - \left( dx^1 \right)^2 + g_{\alpha \beta}
 dx^\alpha dx^\beta
\ee
and the only jumps of  $\Gamma^i_{jk}$
 are
$$
\Gamma^{\pm 1}_{\alpha \beta} = - k^{\pm 1}_{\alpha \beta}
= - \frac{1}{2} g^{\pm }_{\alpha \beta , 1}  \, .
$$
The most natural  definition of $\theta \cdot \delta$  is
 $\frac{1}{2} \delta$
 being equivalent to the choice 
\be
\Gamma^i_{jk} = \frac{1}{2} \left( 
 \Gamma^{+i}_{jk} +  \Gamma^{-i}_{jk}\right)
   \qquad {\rm on} \qquad 
\Sigma \, .
\ee
But cf.  [3] for another choice of  $\Gamma^i_{jk}$   with
 the 
consequence that 
$T^k_{i;k} \ne 0 $ at $\Sigma$. Now the $\delta$ part
 of the equation $T^k_{1;k} =  0 $  reads
\be
\Delta T^1_1 \equiv T^1_{+1} - T^1_{-1} =
  \frac{1}{2} \left(
k_{\alpha \beta }^+ + k_{\alpha \beta }^- \right)
 \tau^{\alpha \beta } \, .
\ee
Analogously one obtains for the other
components
\be
\Delta T^1_\alpha = - \tau^\beta_{\alpha \| \beta}     \, .
\ee
Reintroducing the original 
 coordinate system, the left-hand sides of equations 
(24) and (25) have to be replaced by 
$ \Delta P_1 = \Delta n_i n^k T^i_k$ 
 and 
$\Delta P_\alpha = \Delta n_i e^k_\alpha T^i_k   $, 
respectively; cf. [26].
 Thereby, $\Delta P_i$  
 is the difference of the energy flows on both  sides 
 of  $\Sigma$.

\section*{Acknowledgment}

Discussions with  members of our relativity group, 
especially with  S. Gottl\"ober and U. Kasper, 
are gratefully acknowledged.

\section*{References}

\noindent
1.   Lanczos, K. (1924). Ann. Phys. (Leipzig), {\bf 74} (4), 518.

\noindent
2.   Papapetrou, A., and Treder, H. (1959). Math. Nachr., {\bf 20}, 53.

\noindent
3.   Dautcourt, G. (1963/64). Math. Nachr., {\bf 27}, 277.

\noindent
4.   Israel, W. (1966). Nuovo Cimento B, {\bf 44}, 1.

\noindent
5. Papapetrou, A., Hamoui, A. (1968). Ann. Inst. H. Poincar\'e, {\bf A9}, 179.

\noindent
6.   Kuha\v{r}, K. (1968). Czech. J. Phys., {\bf  B18}, 435.

\noindent
7.   Boulware, D. (1973). Phys Rev. D, {\bf 8}, 2363.

\noindent
8.   Denisov, V. I. (1972). J. Exp. Teor.  Phys., {\bf  62}, 
1990 [Engl. transl (1972) Sov.  Phys. JETP, {\bf 35}, 1038.

\noindent
9.   Lake, K. (1981). Lett. Nuovo Cim. ser. 2, {\bf 30}, 186.

\noindent
10.  Schmidt, H.-J. (1983). Contributed Papers, GR10, 
Padova, 4-9 July 1983, Vol. 1, B. Bertotti, ed. p. 339.

\noindent
11.  Urbandtke, H. (1972). Acta Physica Austriaca, {\bf 35},1.

\noindent
12.  Frehland, E. (1972). Ann. Phys. (Leipzig), {\bf 28}(7),  91.

\noindent
13.  Lake, K. (1979). Phys. Rev. D, {\bf 19}, 2847.

\noindent
14.  Morgan, L., and Morgan, T. (1970). Phys. Rev. D, {\bf 2}, 2756.

\noindent
15.  Voorhees, B. (1972). Phys. Rev. D, {\bf 5}, 2413.

\noindent
16.  Einstein, A. (1939). Ann. Math., {\bf 40},  922.

\noindent 
17.  Gottl\"ober, S. (1980). Diss. A, Akad. Wiss. Berlin.

\noindent
18.  Israel, W., and Stewart, J. (1980). Gen. Rel. Grav. II, A. Held, ed.
 (Plenum, New York), p. 491.

\noindent
 19.    Neugebauer, G. (1980). Relativistische Thermodynamik
 (Akademieverlag, Berlin). 

\noindent
20.  Ono, S., und Kondo, S. (1960). Handbuch der Physik, 
Vol. 10, S. Fl\"ugge, ed. (Springer Verlag, Berlin), p. 134.

\noindent
21.  L\'opes, C. (1981). Nuovo Cimento B, {\bf 66}, 17.

\noindent
22. Krasilnikov, V. A., and Pavlov, V. I.
 (1972). Vestnik  Univ. Moskau (III), {\bf  27}, 235.

\noindent
23. Thomson, W. (Lord Kelvin) (1858). Proc. R. Soc. London, {\bf 9}, 255.

\noindent
24. Taub, A. (1980). J. Math. Phys., {\bf 21}, 1423.

\noindent
25.  Cohen, M., und Cohen, J. (1971). 
Relativity  and Gravitation, C. Kuper, ed. (Gordon/Breach, New
 York), p. 99. 

\noindent
26.  Maeda, K., und Sato, H. (1983). Progr. Theor. Phys., {\bf  70}, 772.

\medskip

\noindent
Received February 14, 1983

\medskip

\noindent 
{\small {In this reprint 
we removed only obvious misprints of the original, which
was published in General Relativity and Gravitation,
Gen. Rel. Grav.  {\bf 16} (1984) Nr. 11, pages 1053 - 1061;  
  Author's address that time:  
Zentralinstitut f\"ur  Astrophysik der Akademie der 
Wissenschaften der DDR, 
1502 Potsdam--Babelsberg, R.-Luxemburg-Str. 17a.}}

\end{document}